\documentclass[preprint,aps,eqsecnum,amssymb,floatfix]{revtex4}
\usepackage{dcolumn}
\usepackage{bm}
\usepackage{graphicx}
\usepackage{color}
\usepackage{morefloats}

\begin{document}
\title{Neutrino-nucleus interactions \\ and the short-range structure of nuclei}

\author{
\vspace{1cm}
F.\ Cavanna$^{\,{\rm a}}$, O.\ Palamara$^{\,{\rm a}}$, R.\ Schiavilla$^{\,{\rm b,c}}$, M. Soderberg$^{\,{\rm a,e}}$, and R.B.\ Wiringa$^{\,{\rm d}}$
\vspace{1cm}
}

\affiliation{
$^{\rm a}$\mbox{Fermi National Acclerator Laboratory, Batavia, IL 60510}\\
$^{\rm b}$\mbox{Thomas Jefferson National Accelerator Facility, Newport News, VA 23606}\\
$^{\rm c}$\mbox{Department of Physics, Old Dominion University, Norfolk, VA 23529}\\
$^{\rm d}$\mbox{Physics Division, Argonne National Laboratory, Argonne, IL 60439}\\
$^{\rm e}$\mbox{Department of Physics, Syracuse University, Syracuse, NY 13244}\\
}
\date{\today}


\maketitle
\section{Introduction}

The atomic nucleus was discovered by Rutherford in 1911~\cite{Rutherford11}, 
who also classified nuclear physics as the {\em unclear physics}.  This 
characterization continues to apply to several areas of nuclear physics 
even today.  The fundamental degrees of freedom in nuclei are believed to be 
quarks and gluons; however, due to color confinement, free quarks are not detectable. 
At low energies, quantum chromodynamics, which governs the behavior of 
interacting quarks and gluons, does not have simple solutions.  The observed
degrees of freedom of nuclear physics are hadrons, protons and neutrons in
particular.  

In the past century many interesting models were developed to 
explain the systematic trends in the low-energy properties of 
stable and near stable atomic nuclei.  They include, for example, 
the liquid-drop model, the compound-nucleus model, the shell model, 
the optical model, the collective model, and the interacting boson 
model.  These models have provided deep insights into nuclear 
structure and reactions, and have been quite successful in correlating many 
of the nuclear properties.  Most of them can be related to the 
shell model which describes the general theory of quantum liquid 
drops.

All the nuclear models tacitly assume that nuclei are made up of interacting
protons and neutrons, collectively referred to as nucleons.  Within this approximation
a general theory can be developed for all low-energy phenomena displayed
by interacting nucleons, ranging from the deuteron to neutron stars.  The simplest
version of this theory describes low-energy nuclear systems as those composed
of nucleons interacting via many-body potentials and many-body electroweak
currents.  We call it the ``basic model.''  It is likely that the shell model and other
models of nuclei can be considered as suitable approximations of this theory
for various energy and mass regions of nuclear systems.

The basic model assumes that a Hamiltonian,
\begin{equation}
\label{nham}
H =  \sum_{i} \left( m_i + \frac{{ \bf p}_i^2}{2\, m_i} \right) +
\sum_{i<j} v_{ij} + \sum_{i<j<k} V_{ijk} + ~...~,
\end{equation}
provides a good approximation to the energy of interacting nucleons.  The
subscripts $i,j,k,...$ label the nucleons in the system.  The mass $m_i$ is that
of a proton or  neutron according to the nature of nucleon $i$.  Much effort
has been devoted over the past several decades on the development  of
nuclear potentials.  Modern two-nucleon ($NN$) potentials $v_{ij}$ consist
of a long-range component induced by one-pion exchange and intermediate- to short-range
components which are modeled either phenomenologically, as in the Argonne $v_{18}$
(AV18) potential~\cite{Wiringa95}, or by scalar and vector meson-exchanges, as
in the CD-Bonn potential~\cite{Machleidt01}, or by a combination of two-pion-exchange
mechanisms and contact two-nucleon terms, such as in chiral-effective-field-theory~\cite{Entem03}. 
All these models fit the extensive $NN$ database ($pp$ and $np$ cross
sections and polarization observables) for energies up to the pion production threshold
with a $\chi^2 \simeq 1$.  However, it is by now an established fact that $NN$ potentials alone
fail to predict the spectra of light nuclei~\cite{Pieper01}, cross sections and analyzing powers
in $Nd$ scattering at low~\cite{Marcucci09} and intermediate~\cite{Kalantar} energies, and
the nuclear matter equilibrium properties~\cite{Akmal}.
\begin{figure}[bth]
\includegraphics[width=6.5in]{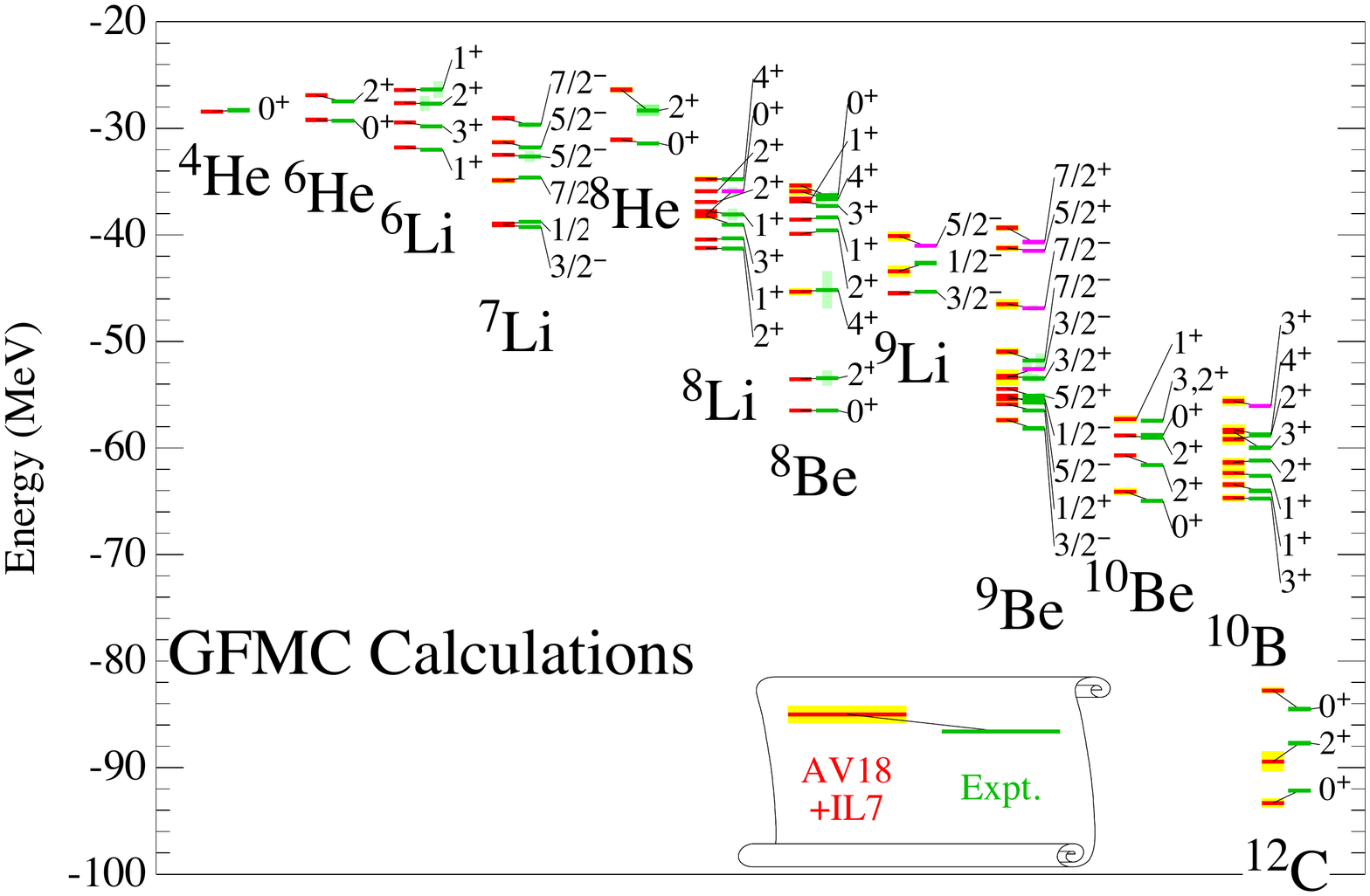}
\caption{\sf \scriptsize The low-lying energy spectra of nuclei with mass number
A=4--12, obtained in exact {\it ab initio} Green's function Monte Carlo (GFMC)
calculations with the AV18/IL7 Hamiltonian.}
\label{fig:f1}
\end{figure}

Models of the three-nucleon potential include two- and three-pion exchange~\cite{IL7,Krebs12}
as well as short-range repulsive terms.  In the Illinois model 7 (IL7), for which the most
extensive calculations have been carried out to date, these multi-pion exchange components
involve excitation of intermediate $\Delta$ resonances.  The IL7 strength is determined by four
parameters which are fixed by a best fit to the energies of about 17 low-lying states of nuclei in the
mass range $ A \leq 10$, obtained in combination with the AV18 $NN$ potential, in exact {\it ab initio}
quantum Monte Carlo calculations.  The resulting AV18/IL7 Hamiltonian then leads to predictions~\cite{Pieper12}
of $\sim 100$ ground- and excited-state energies up to $A$=12, including the $^{12}$C ground-
and Hoyle-state energies, in good agreement with the corresponding empirical values.  Some of
these results are displayed in Fig.~\ref{fig:f1}; in particular, for $^{12}$C  the predicted
ground-state energy and root-mean-square charge radius are~\cite{Lovato13} --93.3(4) MeV and 2.46(2) fm,
respectively, in excellent agreement with the empirical values of --92.16 MeV and 2.471(5) fm.

\section{Short-range structure of nuclei}

The potential $v_{ij}$ that binds protons and neutrons together in
the atomic nucleus is characterized by a strong repulsion
at short ($\lesssim 0.5$ fm) distances, and a strong coupling
between spatial and spin and isospin degrees of freedom
in pairs of nucleons at intermediate to large ($\gtrsim 1$ fm)
separations.  This ``tensor'' character, mostly due to single- and
multi-pion exchanges, of $v_{ij}$ binds the
deuteron---the simplest nucleus consisting of a proton and
neutron---and couples the S- and D-waves into a ground state
with a large non-spherical component.  This is in marked contrast
to systems such as the hydrogen atom where the dominant $1/r$
Coulomb potential leads to a spherical ground state.
\begin{figure}[bth]
\includegraphics[width=4in]{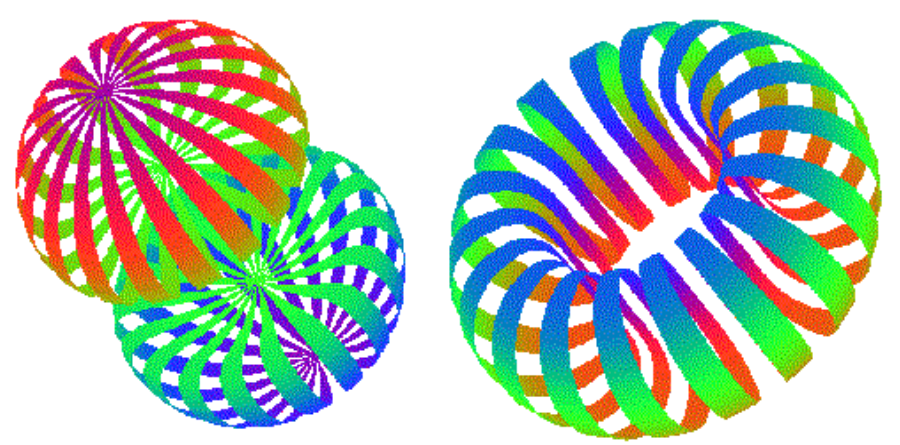}
\caption{\sf \scriptsize Constant density surfaces for a polarized deuteron in the
$M=\pm1$ (left) and $M=0$ (right) states.}
\label{fig:f2}
\end{figure}

While many systems share a trait of short-range repulsion, the tensor
character is unique to the nuclear force.  In the deuteron, which has total
angular momentum $J$=1 and can therefore be oriented in a specific direction,
for example by an external electromagnetic field, with spin projections
$M$=+1 (parallel), $M$=--1 (antiparallel), and $M$=0 (perpendicular),
it leads to the unusual shapes in Fig.~\ref{fig:f2}~\cite{Forest96}---the length of the
dumbbell and the diameter of the torus are both about 1.5 fm.
Their presence has been confirmed in electron scattering
experiments on polarized deuterons.

The correlations induced by the central and tensor components of the
potential $v_{ij}$ affect many nuclear properties.  In nuclei with mass
number $A \ge 3$ the pair density distribution of a neutron and proton
at separation ${\bf r}$ and with their spins coupled to $S=1$ (a deuteron-like
pair) depends strongly on the spin projections $M_S=\pm 1, 0$, i.e., on the relative
orientation between ${\bf r}$ and the spin projection~\cite{Forest96}.  At
a separation $r \simeq 1$ fm the difference in potential energy between
the two configurations in which the neutron and proton are located either along the
$z$-axis (the spin quantization axis) or in the $xy$ plane is large and positive (+200 to
+300 MeV) when $M_S=0$, and large in magnitude but of opposite sign (--150 to
--100 MeV) when $M_S=\pm 1$.  As a consequence of the energy difference between
these two extreme configurations, $np$ pairs in $S=1$ are mostly confined to the $xy$-plane
for $M_S=0$ and mostly along the $z$-axis for $M_S=\pm 1$, giving rise to the peculiar
toroidal- and dumbbell-like equi-density surfaces already shown in Fig.~\ref{fig:f2} for the
deuteron; the holes in the center of the torus and dumbbell are due to the short-range
repulsion.  It turns out~\cite{Forest96} that these $S=1$ $np$-pair densities in different
nuclei can be scaled, through a single scaling factor $R_{Ad}$, to lie on universal
surfaces for $r \lesssim 2$ fm---the dumbbell and torus of the deuteron.  
\begin{figure}[bth]
\includegraphics[width=6.5in]{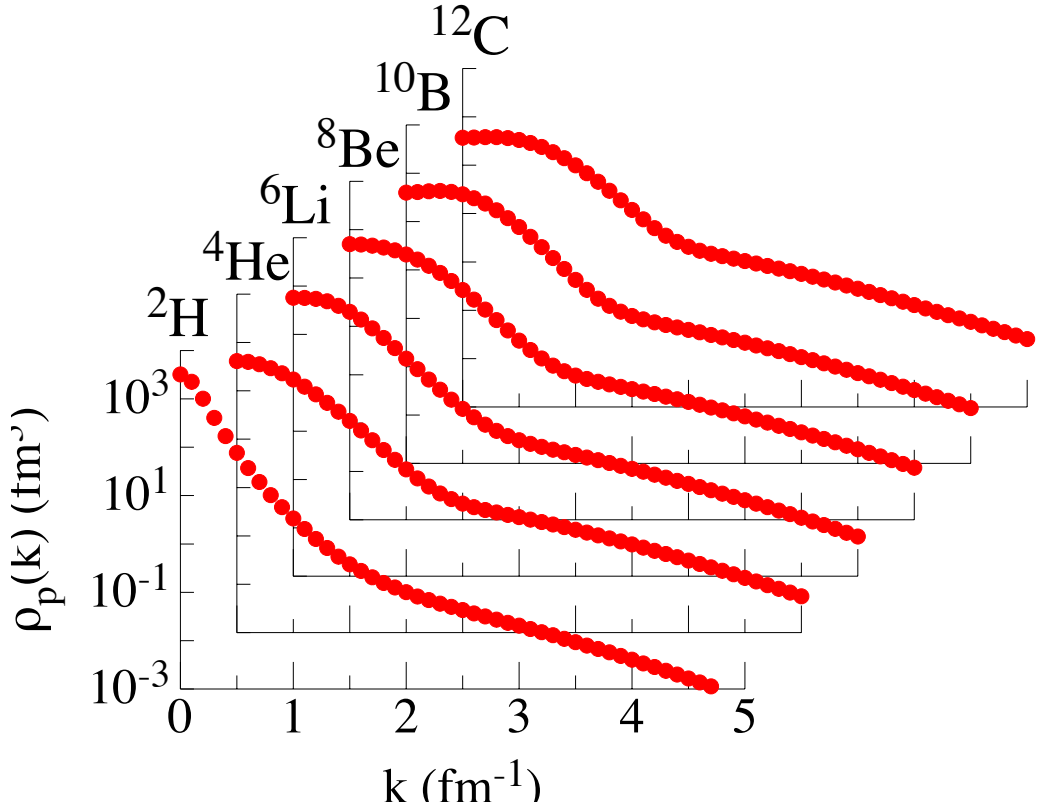}
\caption{\sf \scriptsize Nucleon momentum distributions in s- and p-shell nuclei.}
\label{fig:f3}
\end{figure}

Other nuclear properties strongly affected by correlations are the momentum distributions
of nucleons and nucleon pairs in nuclei, which provide useful insights into various
reactions on nuclei, such as $(e,e^\prime p)$ and $(e,e^\prime pp/np)$ electro-disintegration
processes or neutrino-nucleus interaction experiments.  Single-nucleon momentum distributions
in s- and p-shell nuclei are displayed in Fig.~\ref{fig:f3}~\cite{Wiringa13,Alvioli13}.
Their shape shows a smooth progression as nucleons are added.
As the mass number $A$ increases, the nuclei become more tightly bound,
and the fraction of nucleons at zero momentum decreases.
As nucleons are added to the $p$-shell, the distribution at low momenta
becomes broader, and develops a peak at finite momentum $k$.
The sharp change in slope near $k=2$ fm$^{-1}$ to a broad shoulder is 
present in all these nuclei and is attributable to the strong tensor
correlations induced by the pion-exchange part of the potential $v_{ij}$.
Above $k=4$ fm$^{-1}$, the bulk of the momentum density comes from
correlations associated with the central, spin- and isospin-dependent repulsive
components of $v_{ij}$~\cite{Wiringa13}.

\begin{figure}[bth]
\includegraphics[width=6.5in]{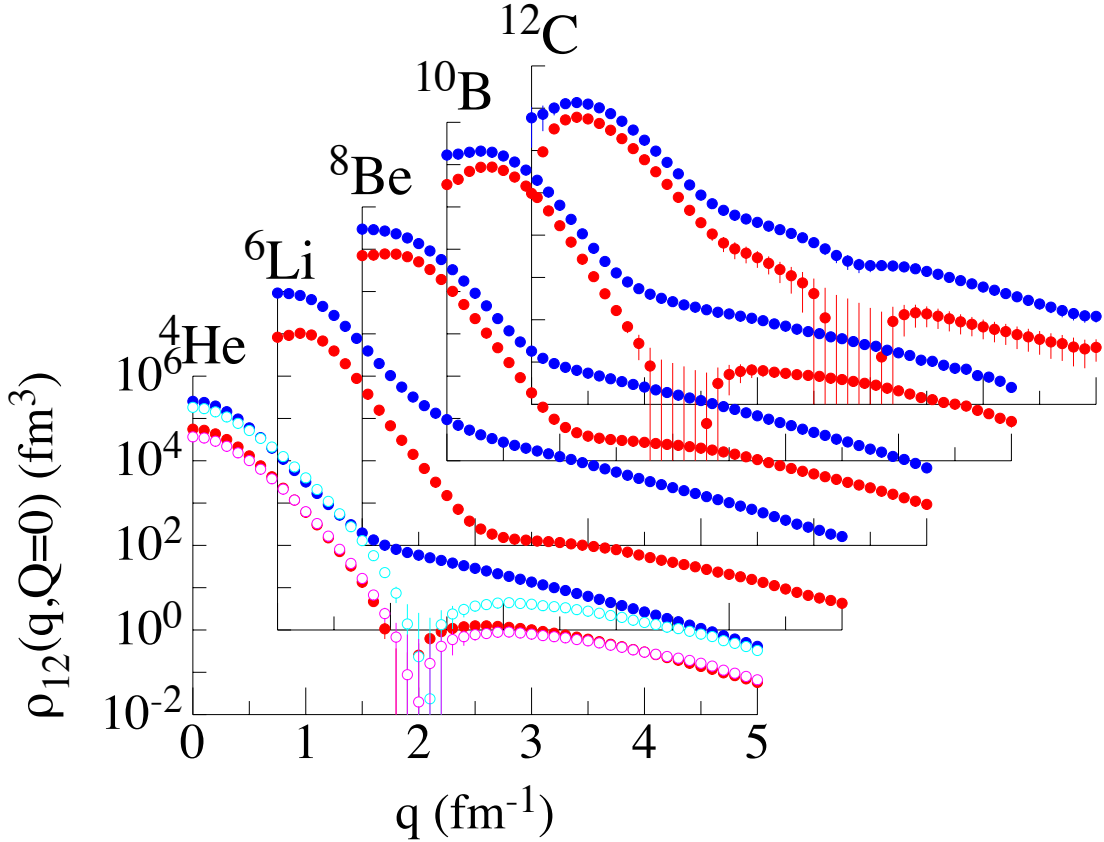}
\caption{\sf \scriptsize The momentum distributions of $np$ and $pp$ pairs in s- and p-shell
nuclei for vanishing total momentum (back-to-back configuration)
and averaged over the directions of the relative momentum ${\bf q}$,
as function of the magnitude $q$.   Also shown for $^4$He is the $np$
momentum distribution corresponding to the AV4$^\prime$ potential
with no tensor component.}
\label{fig:f4}
\end{figure}

The role of the tensor force becomes strikingly apparent when considering
the momentum distributions of back-to-back $np$ and $pp$ pairs in
nuclei~\cite{Schiavilla07} in Fig.~\ref{fig:f4}.  The momentum
distribution of $np$ pairs is much larger than that of $pp$ pairs for relative
momenta in the range of 1.5--3.0 fm$^{-1}$.  The nodal structure present
in the $pp$ momentum distribution is absent in the $np$ one, which instead
exhibits a change in slope at a characteristic value of $q\simeq1.5$ fm$^{-1}$.
In nuclei $pp$ pairs are either in spin $S$=0 and isospin $T$=1 states
or in $ST$=(11) states: the tensor force vanishes in $ST$=(01)
and is weak in $ST$=(11), since the two protons must be in relative
P-wave (or a higher odd partial wave).  On the other hand, most
of the $np$ pairs are in deuteron-like states~\cite{Wiringa06}
for which the tensor force is strongest, since it can act in relative
S-wave.  This is best illustrated by a calculation based on a semi-realistic
potential with only central spin- and isospin-dependent terms but no tensor
term (it is denoted as AV4$^\prime$ below).  This potential
reproduces~\cite{Wiringa02} the empirical S-wave phase shifts
and deuteron binding energy but with only an S-state component; the
D-state, induced by the tensor force, is absent.  While the AV4$^\prime$ $pp$
momentum distribution, shown for $^4$He only in Fig.~\ref{fig:f4}, is similar
to that obtained with the full AV18 except at the lowest
$q$ owing to binding effects, the $np$ momentum distribution develops a node due to the
purely S-wave nature of the deuteron-like state~\cite{Schiavilla07}.

\section{Probing the short-range structure by neutrinos}
Two-nucleon knock-out from high-energy scattering processes is the most appropriate
venue to probe nucleon-nucleon (\textit{NN}) correlations in nuclei.  These
two-nucleon emissions can occur primarily via two mechanisms that will be reviewed hereafter.

The first type of two-nucleon emission mechanism is
one in which the external probe momentum and energy transfers are absorbed
by one of two nucleons in a strongly correlated pair, causing the other (spectator)
nucleon to recoil and be ejected with high momentum~\cite{benhar2}.  More specifically,
in a naive plane-wave-impulse-approximation (PWIA) picture the cross section for such
a process is proportional to the two-nucleon momentum distribution discussed in the
previous section, and is therefore affected by correlations induced by the repulsive core
and tensor character of the \textit{NN} force.

Electron scattering experiments have extensively studied these short-range
correlations (SRC's).  Latest generation experiments have probed them by
triple coincidence reactions of the type $A (e,e^\prime\, np{\rm~or~}pp)A-2$,
in which the two knocked-out nucleons are detected at fixed angles.  The SRC
pair is typically assumed to be at rest prior to scattering and the kinematical
reconstruction utilizes pre-defined four-momentum transfer components
determined from the fixed beam energy and the electron scattering angle and
energy.  The \textit{NN} SRC's are identified by the detection of a pair of
high-momentum nucleons, whose \textit{reconstructed} initial momenta are back-to-back
and exceed the characteristic Fermi momentum of the parent nucleus, while the residual
nucleus is assumed to be left in a highly excited state after the interaction~\cite{Arrington}.
Recent results from JLab (on $^{12}$C) indicate that $\sim 20$\% of the nucleons  (for
$A \ge 12$) act in correlated pairs, and that for high relative momenta ($\sim 400$--500 MeV/c)
$\sim 90$\% of such pairs are in the form of iso-singlet (\textit{np})$_{I=0}$ SRC pairs,
while $\sim 5$\% are in the form of SRC \textit{pp} pairs and, by isospin symmetry, it is
inferred that the remaining $\sim 5$\% are in the form of SRC \textit{nn} pairs~\cite{Science}.
These results are consistent with PWIA expectations: the \textit{np} momentum distribution
is an order of magnitude larger than the \textit{pp} one in back-to-back configurations
and relative momenta in the range of the JLab experiment, see Fig.~\ref{fig:f4}.


The second set of two-nucleon emission mechanisms 
 occur via absorption of the external probe momentum and energy transfers
by a nucleon pair via a two-body mechanism induced by meson exchange (for example, pion-exchange)
or by excitation (de-excitation) of an intermediate nucleon resonance (for
example, the $\Delta$ resonance), followed (preceded) by its de-excitation (excitation)
via meson emission and the consequent absorption of this meson
by a second nucleon~\cite{martini}.  It should be noted that the nucleon-nucleon pairs in these two-body processes may or may not be in a strongly correlated configuration.
 
The latter mechanism seems to be at play in pion absorption in nuclei,
studied extensively over the past several decades in particular 
at intermediate incident energies (i.e., in the $\Delta$-resonance region). 
Pion absorption is highly suppressed on a single nucleon, and hence requires at least a
two-nucleon interaction.  The simplest
mechanism (for $A \ge12$) is on \textit{np} pairs: the so-called quasi-deuteron absorption (QDA),
in which, for example, $\pi^+ + (np)\rightarrow p p$.  Most of the pion energy is carried away
by the ejected nucleons (whose separation energy contributes to the missing energy budget)
and part of the momentum can be transferred to the recoil nucleus (missing momentum).
Observation from bubble-chamber experiments of pairs of energetic protons with three-momentum
$p_{p1},p_{p2} \gtrsim k_F$ ($k_F$ is the Fermi momentum) detected 
at large opening angles in the lab frame (${\rm cos}\,\gamma \lesssim -0.9$) provided 
first hints for SRC's in the target nucleus~\cite{Bellotti}.
 
Neutrino scattering experiments only very recently attempted to explore  SRC's.
The main limitation compared to electron scattering comes from the intrinsic uncertainty
on the four-momentum transfer. 
This originates from the \textit{a priori} undetermined incident neutrino energy.  On the other hand, 
neutrinos (and anti-neutrinos) can effectively probe the nucleus for its SRC content via charge-changing
reactions on SRC pairs leading to {\it two-proton} knock-out topologies.  With the advent of LArTPC detectors
these signatures can be identified directly and unambiguously.  The two protons can indeed be detected
{\it at any emission angle} in the 4$\pi$ sensitive LAr volume and down to energies below the Fermi level
(detection threshold is $T_p^{\rm thr} \simeq 20$ MeV, i.e., well below $k_F \simeq 220$ MeV/c in Ar).
This has been first demonstrated by the ArgoNeuT experiment operated in the NuMI beam (3 GeV average neutrino energy) at Fermilab
in 2009--2010 and complemented by the magnetized MINOS Near-Detector for muon sign and momentum
determination~\cite{ArgoNeuT-2p,ArgoNeuT}. 

Using {\it muon-neutrino} beams, the specific observed reaction is $A(\nu_\mu,\mu^- \,pp)A-2$ reaction,
with the nuclear target $A=^{40}$Ar.  The final state topology consists of a pair of energetic protons
at the interaction vertex accompanying the leading (negative charge) muon.   
\begin{figure}[bth]
\vspace{-0.3cm}
\includegraphics[height=.30\textheight]{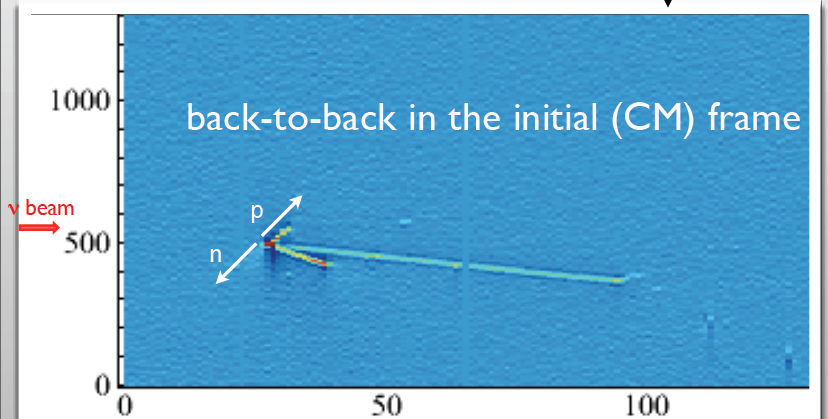}
\caption{\sf \scriptsize Two-dimensional views of one of the events with a reconstructed back-to-back {\it np} pair in the initial state.}
\label{fig:InitState-backToback_event}
\end{figure}
A fraction of the collected events was found compatible with a reconstructed back-to-back configuration of a \textit{np} pair in its {\it initial state} (CdM frame) inside the nucleus,
 a signature compatible with one-body quasi-elastic
interaction on a neutron in a SRC pair~\cite{ArgoNeuT-2p}, much as in the case of the $^{12}$C$(e,e^\prime\, np)$ events observed in the JLab expeirment.
This result was indeed obtained with an approach similar to the electron scattering triple coincidence analysis: the initial momentum of the struck neutron
was determined by transfer-momentum vector subtraction to the higher proton momentum
(${\bf p}_{n}^{i}={\bf p}_{p1} -{\bf q})$ and the lower momentum proton (${\bf p}_{p2}$) was identified
as the recoil spectator nucleon from within SRC, as shown in Fig.~\ref{fig:InitState-backToback_event}.
The momentum transfer ${\bf q}$ is calculated from the reconstructed neutrino energy and the
measured muon kinematics.

\begin{figure}[htbp]
\begin{center}
   \includegraphics[height=.40\textheight]{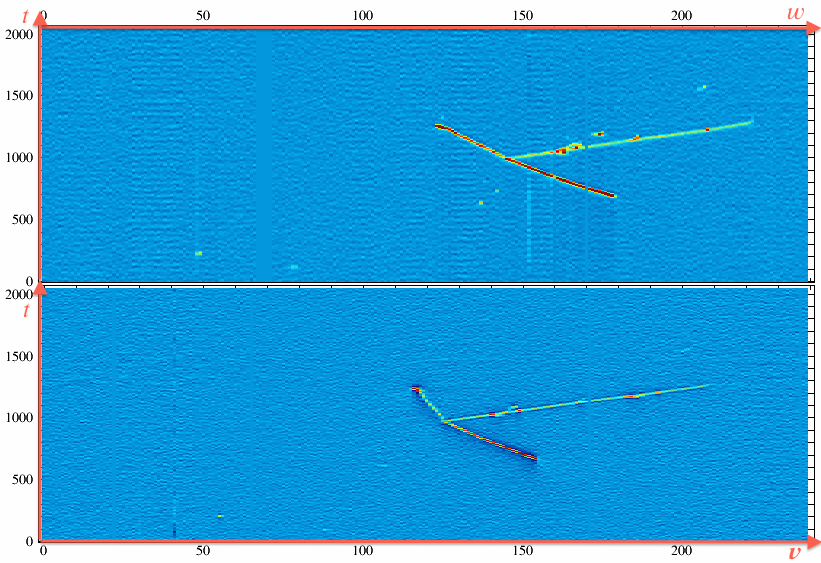}
\caption{\sf \scriptsize Two-dimensional views of one of the four ``hammer events", with a forward
going (negative) muon and a back-to-back proton pair ($p_{p1}=552$ MeV/c and
$p_{p2}=500$ MeV/c).  Transformations from the TPC wire-planes coordinates
(\textit{w,t} ``collection plane" in the top panel; \textit{v,t} ``induction plane" in the
lower panel) into lab coordinates are given in Ref.~\cite{ArgoNeuT}.}
\label{fig:backToback_event_2}
\end{center}
\end{figure}

Another fraction of the ($\mu^-+2p$)
sample detected with ArgoNeuT were found with the two protons in a strictly back-to-back, high momenta
configuration directly observed in the {\it final state} (lab frame)~\cite{ArgoNeuT-2p}.  Visually the signature
of these events gives the appearance of a hammer, with the muon forming the handle and the back-to-back
protons forming the head, see Fig.~\ref{fig:backToback_event_2} for an example of ``hammer" events.

The hammer events are most likely due to pionless (positive charged) resonance mechanisms involving
a pre-existing \textit{np} pair in the nucleus and momentum transfer to the recoil nucleus.

Initial state SRC pairs are nearly at rest, i.e. $\vec {\rm p}_{p}^{~i}\simeq-\vec {\rm p}_{n}^{~i}$.
In the case of CC RES processes with no or low momentum transfer to the pair, the events show a large missing momentum and the two protons in final state have momentum significantly above the Fermi momentum,
 with one almost exactly balanced by the other, i.e. $\vec {\rm p}_{p1}\simeq-\vec {\rm p}_{p2}$.\\
The detection of back-to-back {\it pp} pairs in the Lab frame can thus be seen as ``snapshots" of the initial {\it np} pair configuration.


The number of SRC {\it np} pairs in a nucleus like Ar is large and neutrinos can efficiently detect
them.  The number of SRC {\it pp} pairs is much smaller but would be of great interest to directly probe these structures as well. \\
When switching to a {\it muon-antineutrino} beam, in general, if the $\overline{\nu}_\mu$ is absorbed by $p$ belonging to a SRC $np$, one should produce two back-to-back neutrons with the same frequency as 
 $pp$ with $\nu_\mu$. Their detection however may be problematic from an experimental perspective.
On the other hand, looking again for 
back-to-back {\it two-proton} final state with {\it muon-antineutrino} beam may be a sensitive signature to {\it pp} structures.

\begin{figure}[htbp]
\begin{center}
   \includegraphics[height=.18\textheight]{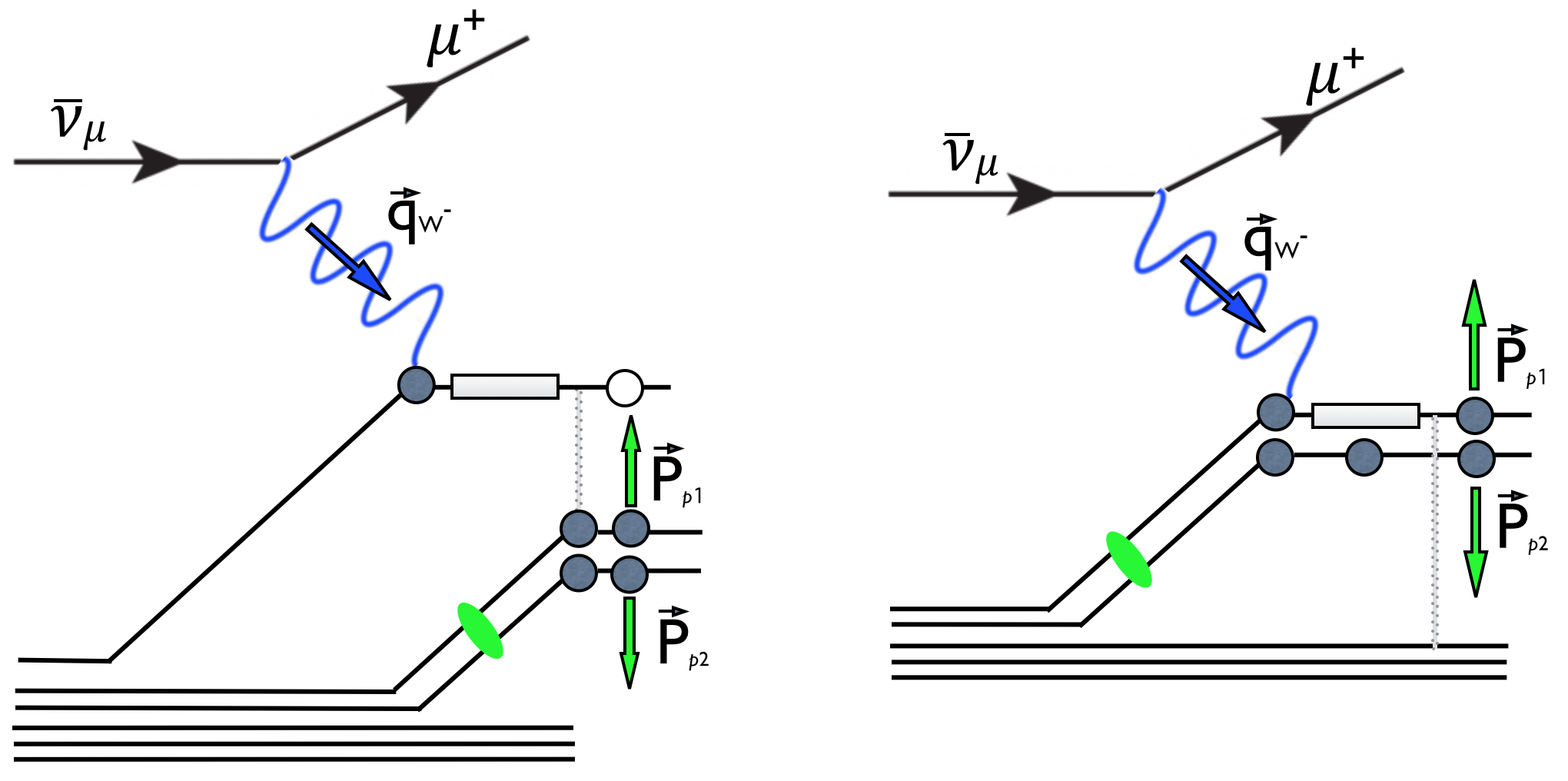}
\caption{\sf \scriptsize Pictorial representations of two-proton knock-out charge-changing reactions involving
\textit{pp} SRC pairs in antineutrino scattering.  Short range correlated nucleon structures in the target
nucleus are denoted by the green symbol, full dots for \textit{p}, open dots for \textit{n}, wide
solid lines (grey) represent neutral nucleon-resonance states, (gray) lines indicate pions.}
\label{Graphs}
\end{center}
\end{figure}
\begin{figure}[htbp]
\begin{center}
   \includegraphics[height=.68\textheight]{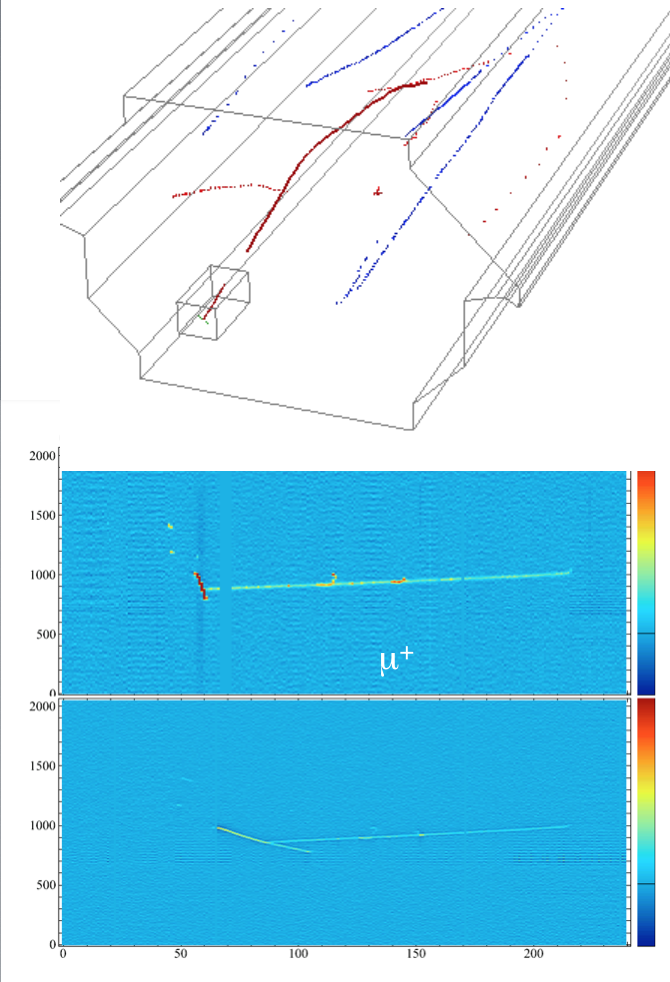}
\caption{\sf \scriptsize Three- and two-dimensional views of an antineutrino ``hammer event", with a forward going positive muon (positive charged particles in MINOS-ND are recognized by curvature in magnetic field, red tracks for positive sign) and a back-to-back proton pair fully contained in the ArgoNeuT TPC active volume.}
\label{mu+}
\end{center}
\end{figure}

The charged current reaction of interest is $A(\bar\nu_\mu,\mu^+\, pp)A-2$,
proceeding as through excitation or de-excitation of a nucleon resonance.  However, in
the anti-neutrino case the formation of a {\it neutral} resonance (e.g., $W^- \, p\rightarrow \Delta^0$)
is required.  One can speculate that resonance-excitation processes involving SRC {\it pp} pairs
of the type shown schematically in Fig.~\ref{Graphs} could contribute.  In these cases two protons
would be knocked-out back-to-back, accompanying the leading positive charged muon,
with a total of {\it three positive charge} particles emitted in final state. 
Indeed, one such event has been found by ArgoNeuT (see Fig.\ref{mu+}) with the two proton tracks fully contained in the active detector volume and the (positive) charge of the muon unambiguously determined
in the downstream MINOS-ND detector.   

The detection of these types of events (back-to-back $2p+\mu^-$ and $2p+\mu^+$
shows that mechanisms directly involving SRC pairs ({\it np} and {\it pp}) in the nucleus
are active and can be efficiently explored in neutrino-Ar and antineutrino-Ar interactions with the LAr TPC technology.
The event statistics from ArgoNeuT is very limited and cannot provide definitive conclusions. However, larger mass and high statistics LAr-TPC detectors 
have the opportunity to clarify the issue in the near future. The MicroBooNE experiment with about 70 t of active LAr mass (compared to ArgoNeuT's $<$0.5 t) is expected to begin operations on the Booster neutrino beam 
(0.8 GeV average energy) at Fermilab in early 2015. MicroBooNE will be capable of performing a systematic study of SRC (as well as of yet unexplored nuclear effects) at an unprecedented level of detail and statistics. 
The inclusion of a realistic and exhaustive treatment of SRC in the one- and two-body component of the nuclear current in current theoretical modeling is thus necessary and urgent.  The subsequent MC implementation is also extremely important for comparison with liquid argon data.  

Improvements in theoretical modeling and MC treatment of these short range phenomena, and comparisons with data, will require sustained collaboration between nuclear theorists and neutrino experimentalists.  The extensive history of studying this area of nuclear physics in electron- and hadron-scattering experiments, coupled with the transformative capabilities of LArTPCs to identify neutrinos, will provide collaborators in this endeavor a ripe opportunity for new discoveries that will further our understanding of the nucleus.
%
%
%
%

%
%
\end{document}